\documentclass[runningheads,orivec,a4paper]{llncs}
\usepackage{graphicx}
\usepackage{hyperref}
\usepackage{amssymb}
\usepackage{tikz}
\usepackage{amsmath}
\tikzstyle{par}=[draw,circle,above,thick,inner sep=1.5pt,minimum height=1pt,fill=gray]
\tikzstyle{var}=[draw,circle,above,thick,minimum height=.5cm]
\tikzstyle{varobs}=[draw,circle,above,thick,minimum height=.5cm,fill=gray!30]
\tikzstyle{plate}=[draw,rectangle,rounded corners=3pt,thick]
\titlerunning{Hetero-Modal Variational Encoder-Decoder}
\newcommand{\figref}[1]{Fig.~\ref{#1}}
\newcommand{\Figref}[1]{Figure~\ref{#1}}
\newcommand{\tabref}[1]{Table~\ref{#1}}

\DeclareMathOperator{\ELBO}{ELBO}
\DeclareMathOperator{\KL}{KL}

\usepackage{xcolor,colortbl}

\definecolor{Gray}{gray}{0.85}
\definecolor{LightGray}{gray}{0.95}

\newcolumntype{a}{>{\columncolor{Gray}}c}
\newcolumntype{b}{>{\columncolor{white}}c}

\begin{document}
\authorrunning{R. Dorent et al.}
\title{Hetero-Modal Variational Encoder-Decoder for Joint Modality Completion and Segmentation}
%
%
\author{Reuben Dorent, Samuel Joutard, Marc Modat, S\'ebastien Ourselin and Tom Vercauteren}
%
%
\institute{School of Biomedical Engineering and Imaging Sciences, King’s College London
\email{reuben.dorent@kcl.ac.uk}\\}
\maketitle              
\begin{abstract}
We propose a new deep learning method for tumour segmentation when dealing with missing imaging modalities. Instead of producing one network for each possible subset of observed modalities or using arithmetic operations to combine feature maps, our hetero-modal variational 3D encoder-decoder independently embeds all observed modalities into a shared latent representation. Missing data and tumour segmentation can be then generated from this embedding. In our scenario, the input is a random subset of modalities.  We demonstrate that the optimisation problem can be seen as a mixture sampling. In addition to this, we introduce a new network architecture building upon both the 3D U-Net and the Multi-Modal Variational Auto-Encoder (MVAE). Finally, we evaluate our method on BraTS2018 using subsets of the imaging modalities as input. Our model outperforms the current state-of-the-art method for dealing with missing modalities and achieves similar performance to the subset-specific equivalent networks.
\end{abstract}
\section{Introduction}
Tumour segmentation and associated volume quantification plays an essential role during the diagnosis, follow-up and surgical planning stages of primary brain tumours. Multiple imaging sequences are usually employed to distinguish and assess the key tumour components such as the whole tumour, the peritumoral edema and the enhancing region. The common sequences are T1-weighted (T1), contrast enhanced T1-weighted (T1c), T2-weighted (T2) and Fluid Attenuation Inversion Recovery (FLAIR) images. These modalities reveal different characteristics of brain tissues. In practice, the set of acquired modalities may vary during the clinical assessment. For this reason, we aim to automatically segment these key components given an arbitrary set of modalities. 

Methods based on deep learning currently achieve the best performance in brain tumour segmentation. Most of them require the full set of $n$ modalities as input \cite{nonewnet,nvidianet}, while a scenario of missing modalities is common in practice. Segmentation with missing data can be achieved by: 1/~Training a model for each possible subset of modalities; 2/~Synthesising missing modalities \cite{completion}
in order to then perform full modality segmentation; 3/~Creating a common feature space which encodes the shared information from which the segmentation is created \cite{hemis,pimms}. The two first options involve training and handling a different network for each of the $2^{n}-1$ combinations.
These two solutions are cumbersome and computationally sub-optimal since duplicate information is extracted $2^{n}-1$ times. In contrast, encoding the modalities into a common feature space produces a single model that shares feature extraction.

The current state-of-the-art network architecture which allows for missing modalities is HeMIS \cite{hemis} and
related extensions~\cite{pimms}. Feature maps are first extracted independently for each modality, then their first and second moments are computed across the modalities and used for predicting the final segmentation. However, using these arithmetic operations does not force the network to learn a shared latent representation. In contrast, Multi-modal Variational Auto-Encoders (MVAE) \cite{mvae} provide a principled formulation to create a common representation: the $n$ modalities and the segmentation map are considered conditionally independent given the common latent variable $z$.

While our goal to segment the tumour with missing modalities, auto-encoding and modality completion promote informativeness of the latent space and can be seen as regularizers, similarly to \cite{nvidianet}. Ideally, all the modality-specific information should be encoded in the common latent space, meaning that the model should be able to reconstruct all the observed modalities. Additionally, the information loss related to any missing modality should be minimal (modality completion). 

In this paper, we introduce a hetero-modal variational encoder-decoder for tumour segmentation and missing modalities completion. The contribution of this work is four-fold. First, we extend the MVAE for 3D tumour segmentation from multimodal datasets with missing modalities.
Secondly, we propose a principled formulation of the optimisation process based on a mixture sampling procedure. Thirdly, we adapt the 3D U-Net in a variational framework
for this task.
Finally, we show that our model outperforms HeMIS in terms of tumour segmentation while comparing favourably with equivalent subset-specific models.

\section{Method}
\subsection{Multi-modal Variational Auto-Encoders (MVAE)}
The MVAE~\cite{mvae} aims at identifying a model in which $n$ modalities $\textbf{x}=(x_1,..,x_n)$ are conditionally independent given a hidden latent variable $z$. We consider the directed latent-variable model parameterised by $\theta$ (typically the weights of a decoding network $f_{\theta}(\cdot)$ going from the latent space to the image space):
\begin{equation}
    \label{graphical_model}
    p_{\theta}(z,x_1,...,x_n) = p(z) \prod_{i=1}^n p_{\theta}(x_i|z)
\end{equation}
where $p(z)$ is a prior on the latent space, which we classically choose as a standard normal distribution $z \sim \mathcal{N}(0,I)$. 
The goal is then to maximise the marginal log-likelihood $\mathcal{L}(\mathbf{x};\theta)=\log(p_{\theta}(x_1,...,x_n))$ with respect to $\theta$. However, the integral $p_{\theta}(x_1,...,x_n)=\int p_{\theta}(\mathbf{x}|z)p(z)$  is computationally intractable. \cite{vae} proposed to optimise, with respect to $(\phi,\theta)$, the evidence lower-bound (ELBO):
\begin{equation}
    \label{eqq:elbo_vae}
    \mathcal{L}(\mathbf{x};\theta) \geq \ELBO(\mathbf{x};\theta,\phi) \triangleq E_{q_{\phi}(z|\mathbf{x})}[\log(p_{\theta}(\mathbf{x}|z))] - \KL[q_{\phi}(z|\mathbf{x})||p(z)]
\end{equation}
where $q_{\phi}(z|\mathbf{x})$ is a tractable variational posterior that aims to approximate the intractable true posterior $p_{\theta}(z|\mathbf{x})$.
For this purpose, $q_{\phi}(z|\mathbf{x})$ is typically modelled as a Gaussian after an encoding of $\mathbf{x}$ into a mean and diagonal covariance by a neural network, $h_{\phi}(\mathbf{x})=\big(\mu_{\phi}(\mathbf{x}),\Sigma_{\phi}(\mathbf{x})\big)$, such that:
\begin{equation}
    \label{vae_eq}
    q_{\phi}(z|\mathbf{x}) = \mathcal{N}(z; \mu_{\phi}(\mathbf{x}), \Sigma_{\phi}(\mathbf{x}))
\end{equation}
 The KL divergence between the two Gaussians $q_{\phi}(z|\mathbf{x})$ and $p(z)$ can be computed in closed form given by their means and covariances.
 In contrast, estimating $E_{q_{\phi}(z|\mathbf{x})}[\log(p_{\theta}(\mathbf{x}|z))]$ is done by sampling the hidden variable $z$ according to the Gaussian $q_{\phi}(\cdot|\mathbf{x})$ and then decoding it as $f_{\theta}(z)$ in image space to evaluate $p_{\theta}(\mathbf{x}|z)$. 
%
To make sampling from $z|\mathbf{x}$ amenable to back-propagation, reparametrisation is used~\cite{vae}: $\mu_{\phi}(\mathbf{x}) + \Sigma_{\phi}(\mathbf{x}) \times \epsilon $ where $\epsilon \sim \mathcal{N}(0,I)$.

Wu, \textit{et al.} \cite{mvae} extended this variational formulation to a multi-modal setting. The authors remarked that $p_{\theta}(z|\mathbf{x}) \propto p(z)\prod_{i=1}^n\frac{p_{\theta}(z|x_i)}{p(z)}$. This expression shows that $p_{\theta}(z|\mathbf{x})$ can be decomposed into $n$ modality-specific terms. For this reason, the authors approximate each $\frac{p_{\theta}(z|x_i)}{p(z)}$ with a modality-specific variational posterior $q_{\phi_i}(z|x_i)$. Similarly to \eqref{vae_eq}, $q_{\phi_i}(z|x_i)$ is modelled as a Gaussian distribution after an encoding of $x_i$ into a mean and a diagonal covariance by a neural network, $h_{\phi_i}(x_i)=\big(\mu_{\phi_i}(x_i),\Sigma_{\phi_i}(x_i)\big)$, such that $q_{i}(z|x_i)= \mathcal{N}(z;\mu_{\phi_i}(x_i),\Sigma_{\phi_i}(x_i))$.  Finally, \cite{gaussian_product} demonstrates that $q_{\phi}(z|\mathbf{x}) \propto p(z) \prod_{i=1}^n q_{\phi_i}(z|x_i)$ is Gaussian with mean $\mu_{\phi}$ and covariance $\Sigma_{\phi}$ defined by:
\begin{equation}
    \Sigma_{\phi}=(I + \sum_i \Sigma_{\phi_i}^{-1})^{-1}
    \text{ and }
    \mu_{\phi} = \Sigma_{\phi}^{-1} (\sum_i \Sigma_{\phi_i}^{-1} \mu_{\phi_i})  
    \label{product_gaussian_formula}
\end{equation} 
This formulation allows for encoding each modality independently and fusing their encoding using a closed-form formula.
%

However, from this well-posed multimodal extension of the ELBO, \cite{mvae} resort to a ad hoc training sampling procedure. At each training iteration, the extremes cases (one modality and all the modalities) and random modality subsets are used concurrently. This option is highly memory consuming, not suitable for 3D images and not adapted to the clinical scenarios where some imaging subsets are clinically more frequent than others. The next section proposes to include this prior information in our principled training procedure via ancestral sampling.

\subsection{Mixture Sampling for Modality Completion and Segmentation}
\begin{figure}[tb!]
\begin{center}
\includegraphics[width=0.9\linewidth]{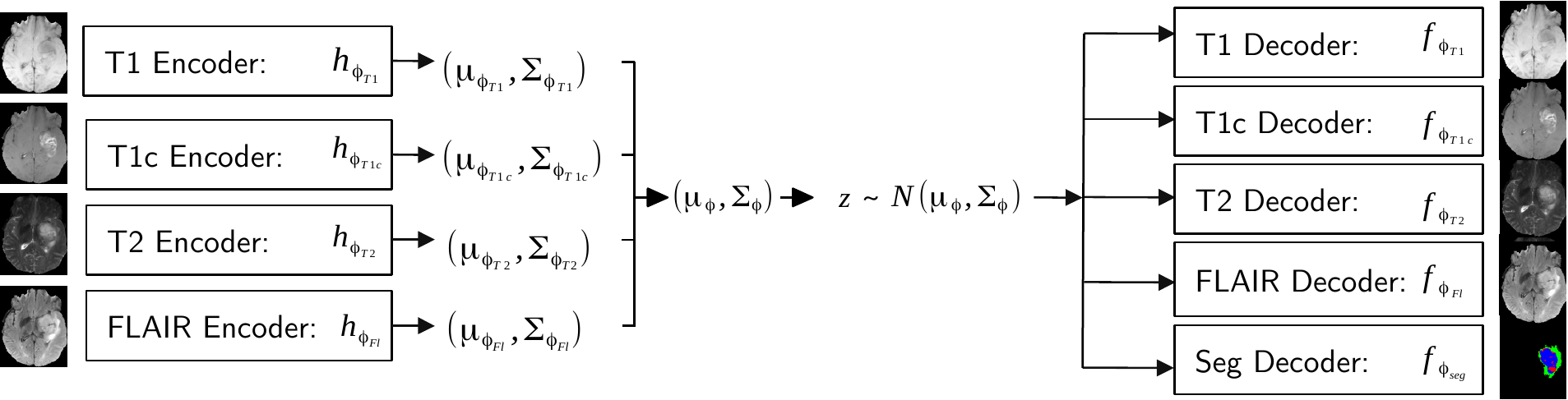}
\caption{MVAE architecture. Each imaging modality is encoded independently, the mean and covariance of each $q(z|x_i)$ are fused using the closed-form formula \eqref{product_gaussian_formula}. A sample $z$ is randomly drawn and is decoded into imaging modalities and the segmentation map.}
\label{graph}
\end{center}
\end{figure}

In our scenario, the clinician provides a subset of $n=4$ imaging modalities with some subsets of input modalities being more likely to be provided than others.
We use an encoder-decoder to produce the missing modalities as well as the tumour segmentation.
Although segmentation could be considered as a missing modality, we chose not to encode it as it is not observed in practice. Consequently, our model is composed of 4 encoders and 5 decoders (see \figref{graph}).

Without loss of generality, we consider a training set providing the complete $n$ modalities per subject. Consequently, during training, we can artificially remove some modalities as input yet evaluate the reconstruction error on all the modalities. When the training set is incomplete, the reconstruction error is only evaluated on the available data.


Let $\mathcal{P}$ denote the set of all possible non-empty combinations of the $n$ modalities. Our goal is to maximise 
\eqref{eqq:elbo_vae}
when $z$ has been encoded via a random subset ${\pi}\in\mathcal{P}$ drawn with probability $\alpha_{\pi}$. This is exactly the ancestral sampling of a mixture model: we first draw the class label (here the subset) and then we draw a sample from the distribution associated to this class. For this reason, we model $q_{\phi}(z|\mathbf{x})$ as a mixture where the probabilities $\alpha_{\pi}$ are chosen to be representative of the clinical scenario:
$$q_{\phi}(z|\mathbf{x}) = \sum_{\pi \in \mathcal{P}} \alpha_{\pi} q_{\phi}^{\pi}(z|\mathbf{x_{\pi}})$$
%
%
%
We choose $q_{\phi}^{\pi}(z|\mathbf{x_{\pi}})$ as Gaussian.
%
%
Given the convexity of the KL divergence and the fact that $\sum_{\pi \in \mathcal{P}}\alpha_{\pi}=1$, we obtain:
$$ \KL[q_{\phi}(z|\mathbf{x})||p(z)] \leq  \sum_{\pi}\alpha_{\pi}\KL[q_{\phi}^{\pi}(z|\mathbf{x_{\pi}})||p(z)]$$
Finally, our lower-bound is a weighted sum of the subset-specific lower-bound:
\begin{equation}\label{eq:elbo_mvae}
\mathcal{L}(\mathbf{x};\theta) \geq \sum_{\pi \in \mathcal{P}} \alpha_{\pi}(\underbrace{E_{q_{\phi}^{\pi}(z|\mathbf{x_\pi})}[\log(p_{\theta}(\mathbf{x}|z))] - \KL[q_{\phi}^{\pi}(z|\mathbf{x_{\pi}})||p(z)]}_{\ELBO_{\pi}(\mathbf{x})})
\end{equation}
The single Gaussian prior model for $p(z)$ promotes consistency of the embedding $z$ across the subsets of modalities $\pi$ ($q_{\phi}^{\pi}(z|\mathbf{x_{\pi}})$) and in turn across the full set of modalities ($q_{\phi}(z|\mathbf{x})$).
In our optimisation procedure, at each iteration, we propose to randomly draw a subset $\pi$ with a probability $\alpha_{\pi}$ as the model input and optimise $\ELBO_{\pi}(\mathbf{x})$.
Classical modelling of $p_{\theta}(.|z)$ includes
Gaussian distribution for image reconstruction and Bernoulli distribution for classification.


\subsection{Network architecture: 3D Variational Encoder-Decoder}
To exploit our framework we propose a novel network architecture: a 3D encoder-decoder with variational skip-connections. Our model is a mix between a 3D U-Net \cite{unet} and the MVAE \cite{mvae}.\

In the U-net architecture, context information is extracted via the contracting path (encoder) and precise localisation is produced by the expanding part (decoder). In addition, information is captured at different levels via the skip-connections. To avoid a trivial identity function, existing auto-encoder architectures do not use skip-connections. In our case, the encoding of the latent variable is multi-modal and 
the imposed consistency of the latent representation creates a bottleneck.
Skip-connections therefore do not allow for trivial identity mapping and can be included in our architecture. 
\begin{figure}[tb!]
\includegraphics[width=\textwidth]{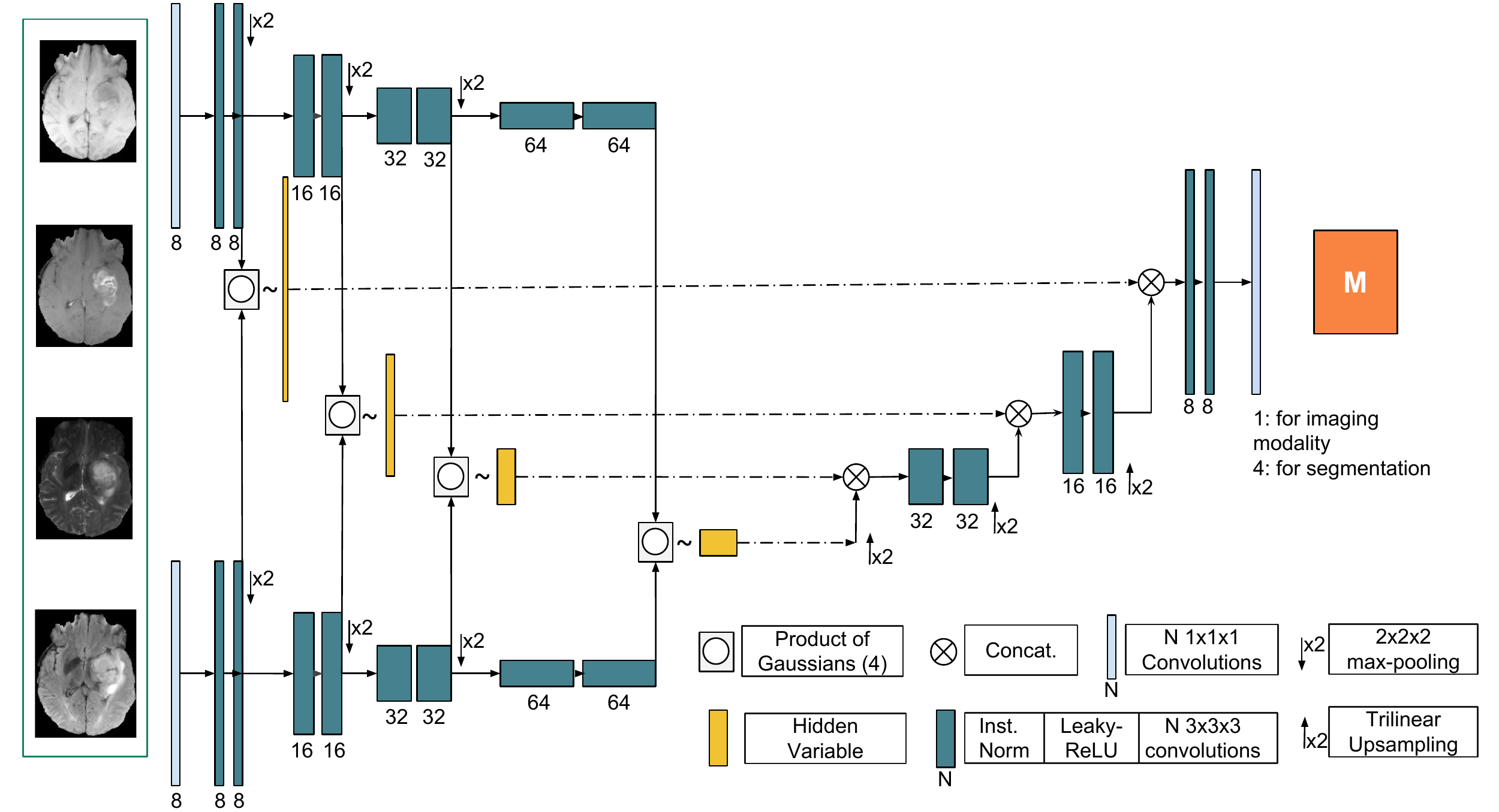}
\caption{Our 3D variational encoder-decoder (\textit{U-HVED}). Only two encoders and one decoders are shown. Product of Gaussian is defined in \eqref{product_gaussian_formula}.} \label{network}
\end{figure}
%
%

We propose to use a multi-level latent variable to generate them. \Figref{network} shows our network architecture. Unlike the existing hierarchical VAE models \cite{NIPS2016_6275,ICML_hiearchical}, we propose a fully convolutional network. Each modality $i$ is independently encoded which produces 4 multi-scale means and variances $(\mu_{i}^{k},\Sigma_{i}^{k})_{k\in[1,..,4]}$. At each level, the means and the variances of the modalities present in the input subset $x_{\pi}$ are combined via the product of Gaussian defined in \eqref{product_gaussian_formula}. We then decode the multi-scale latent variable for each of the modalities and the segmentation. Consequently, we have $n$ encoders and $n+1$ decoders. We assert that it is the first deep network which allows for missing modalities and performs 3D imaging reconstruction and segmentation in a variational manner.

\section{Data and implementation details.}
\subsubsection{Data.}
We evaluate our method on the training set of BRATS18 \cite{brats}. The training set contains the scans of 285 patients, 210 with high grade glioma and 75 with low grade glioma. Each patient was scanned with four sequences (T1, T1c, T2 and FLAIR) and pre-processed by the organisers: scans have been skull-striped and re-sampled to an isotropic 1mm resolution, and the four sequences of the each patient have been co-registered. The ground truth was obtained by manual segmentation results given by experts. The  segmentation classes include the following tumour tissue labels: 1) necrotic core and non-enhancing tumour, 2) oedema, 3) enhancing core. 

\subsubsection{Implementation details.}
As pre-processing step, we used histogram-based scale standardisation method \cite{histo} followed by a zero mean and unit-variance normalisation. As a data augmentation, we randomly flip the axes and include a rotation with a random angle in $[-10^{\circ},10^{\circ}]$. The networks were implemented in Tensorflow using NiftyNet \cite{niftynet}. We used Adam as optimiser with initial learning rate $10^{-3}$ divided by 4 every $10^4$ iterations, batch size 1 and maximal iteration 60k. Early stopping is performed if a plateau of performance is reached on the validation data set. At each iteration, a $112\times112\times112$ random patch is fed to the network. We did a 3-fold validation by random split of the data set a training ($70\%$), validation ($10\%$) and testing ($20\%$) sets. We regularize with a $L2$ weight decay of $10^{-5}$.
During training, we uniformly draw a number of modalities $i$ between 1 and 4 and uniformly draw a subset $\pi$ of size $i$. During inference, given a subset of modalities, we randomly draw 10 hidden variable $z$ from $q(.|\mathbf{x_{\pi}})$ and decode them and average the outputs. Implementation is publicly available\footnote{\url{https://github.com/ReubenDo/U-HVED}}.

\subsubsection{Choices of the losses.}
The reconstruction loss follows from $p_{\theta}(x_i|z)$. For the segmentation we use the sum of the cross-entropy $L_{cross}$ and the dice loss function $L_{dice}$ \cite{nonewnet}. For the imaging reconstruction loss, we used the classic $L_{2}$ loss. Additionally, given a drawn subset $\pi$, our loss includes the closed-form KL divergence between the Gaussians $q_{\phi}(z|\mathbf{x_{\pi}})$ and $p(z)$. For weighting the regularization losses (KL divergence and reconstruction loss), we did a grid search over weights in $[0, 0.1, 1]$. Finally, the loss associated to maximising the ELBO \eqref{eq:elbo_mvae} is:
$$L=L_{dice} + L_{cross} + 0.1*L_2 + 0.1*\KL$$

\section{Experiments and results}
\subsubsection{Model comparison.}
To evaluate the performance of our model (\textit{U-HVED}), we compare it to three different approaches:
The first, \textit{HeMIS} is the model described in \cite{hemis} and is the current state-of-the-art for segmentation with missing modalities. The second, \textit{U-HeMIS}, is a particular case of our method where the modalities are encoded as \textit{U-HVED} and the skip-connection are the first and second moments of the modality-specific feature maps such as in \textit{HeMIS}. \textit{U-HeMIS} has only one decoder for tumour segmentation. The third approach, \textit{Single}, is the "brute-force" method in which for each possible subset of modalities, we train a U-Net network where the observed modalities are concatenated as input. The encoder and decoder are those of our model. Given the 3-fold validation, we consequently trained 45 \textit{Single} networks.

\subsubsection{Missing modalities completion.}
\begin{figure}[tp!]
\begin{center}
\includegraphics[width=0.9\textwidth]{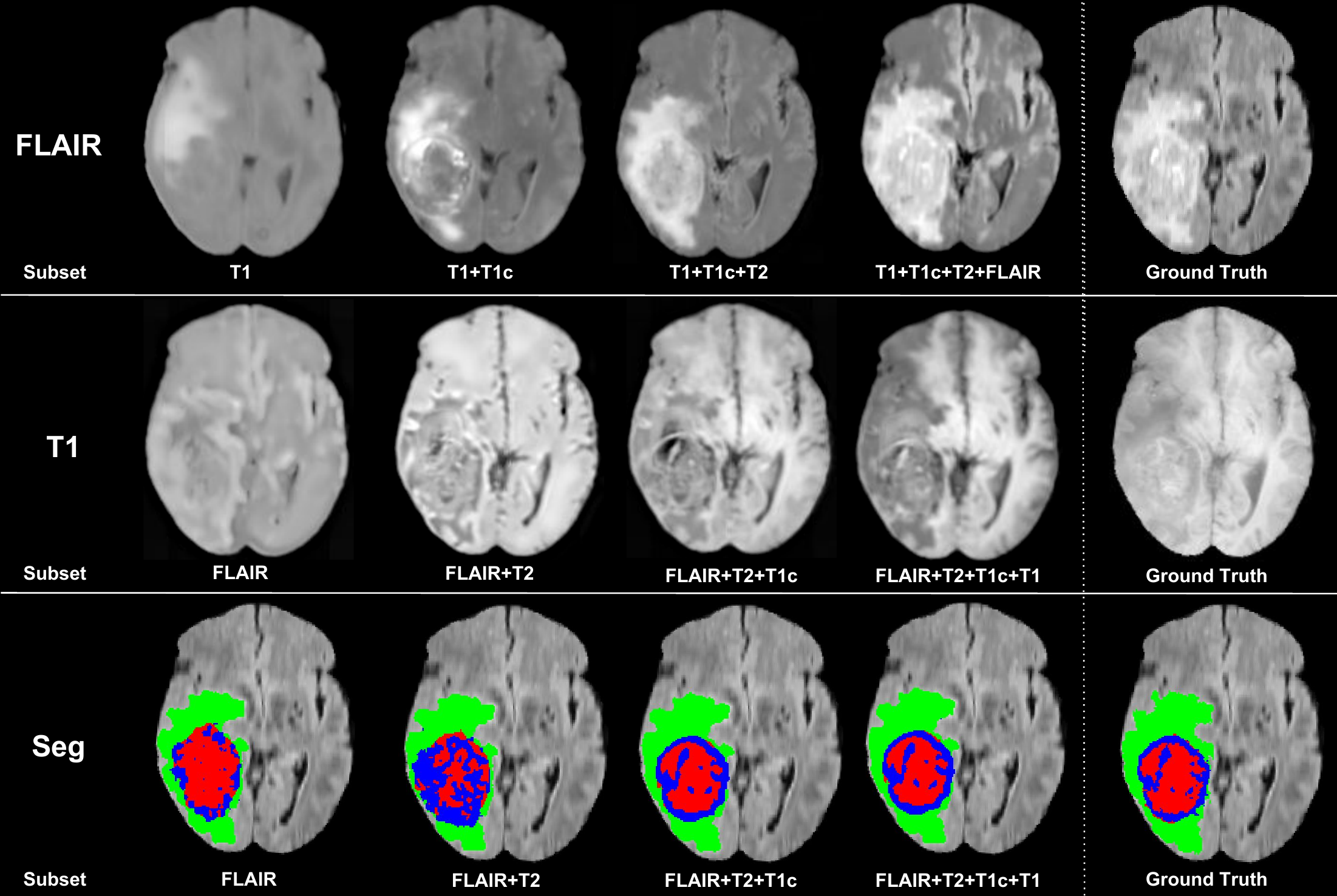}
\caption{Example of FLAIR and T1 completion and tumour segmentation given a subset of modalities as input. Green: edema; Red: non-enhancing core; Blue: enhancing core.} \label{missing_seg}
\end{center}
\end{figure}

Unlike these three approaches, \textit{U-HVED} (\textit{Ours}) generates missing modalities. Since image completion is a means rather than an end, we only provided a qualitative evaluation (\figref{missing_seg}) of T1 and FLAIR reconstruction examples. We find the reconstruction to be good quality, given that VAEs classically suffer of blurriness. Interestingly, our model tries to reconstruct the tumour information even when the tumour information is missing or not clear, such as in T1 scans. Moreover, comparable reconstructions are performed using 3 modalities and 4 modalities. This suggests that our network can effectively learn a common representation of the imaging modalities.

\subsubsection{Tumour segmentation.}
In order to evaluate the robustness of our model, we present qualitative results in \figref{missing_seg} and comparative results with other methods in \tabref{table_scores} for all the possible input subsets. We used the Dice Similarity as metric. 
First, the U-Net architecture in \textit{U-HeMIS} always achieves better performance than the original  2D fully-convolutionnal \textit{HeMIS}. This highlights the efficiency of the 3D U-net architecture. Secondly, \textit{U-HVED} (\textit{Ours}) outperforms significantly \textit{U-HeMIS} in most of the cases: 13 out of 15 cases for the complete tumour, 10 out of 15 cases for the core tumour; 11 out 15 cases for the enhancing tumour. This demonstrates that auto-encoding and modality completion improves the segmentation performance. Finally, \textit{U-HVED} achieves similar performance to the 15 subset-specific models (\textit{Single}). Again, this suggests that the imaging modalities are efficiently embedded in the latent space. 

\begin{table}[tb!]
\caption{Comparison of the different models (Dice $\%$) for the different combinations of available modalities. Modalities present are denoted by $\bullet$, the missing ones by $\circ$. $^{*}$ denotes significant improvement provided by a Wilcoxon test ($p < 0.05$).}\label{table_scores}
\resizebox{\textwidth}{!}{
\begin{tabular}{cccc|ccc|c||ccc|c||ccc|c}
\hline
\multicolumn{4}{c|}{Modalities}                                                                                                                & \multicolumn{4}{c||}{Complete}                                    & \multicolumn{4}{c||}{Core}                                        & \multicolumn{4}{c}{Enhancing}                                                               \\ \hline
\rowcolor{Gray}

{$F$} & {$T_1$} & {$T_1c$} & {$T_2$} & {HeMIS} & {U-HeMIS} & {U-HVED} & {Sing} & {HeMIS} & {U-HeMIS} & {U-HVED} & {Sing} & {HeMIS} & {U-HeMIS} & {U-HVED} & {Sing} \\ \hline
{$\circ$}  & {$\circ$}  & {$\circ$}  & {$\bullet$} & 38.6 & 79.2 & $\textbf{80.9}^{*}$ & 82.6 & 19.5 & 50.0 & $\textbf{54.1}^{*}$ & 54.9 & 0.0 & 23.3 & $\textbf{30.8}^{*}$ & 34.2 \\ \rowcolor{LightGray}

{$\circ$}  & {$\circ$}  & {$\bullet$} & {$\circ$}  &  2.6  & 58.5 & $\textbf{62.4}^{*}$ & 70.4 & 6.5 & 58.5 & $\textbf{66.7}^{*}$ & 71.5 & 11.1 & 60.8 & $\textbf{65.5}^{*}$ & 70.4\\

{$\circ$}  & {$\bullet$} & {$\circ$}  & {$\circ$}  & 0.0 & $\textbf{54.3}^{*}$ & 52.4 & 72.7 & 0.0 & $\textbf{37.9}$ & 37.2 & 59.2 & 0.0 & 12.4 & $\textbf{13.7}^{*}$ & 32.2\\ \rowcolor{LightGray}

{$\bullet$} & {$\circ$}  & {$\circ$}  & {$\circ$}  &  55.2 & 79.9 & $\textbf{82.1}^{*}$ & 81.5 & 16.2 & 49.8 & $\textbf{50.4}$ & 55.5 & 6.6 & $\textbf{24.9}$ & 24.8 & 26.3 \\

{$\circ$}  & {$\circ$}  & {$\bullet$} & {$\bullet$} &48.2 & 81.0 & $\textbf{82.7}^{*}$ & 83.2 & 45.8 & 69.1 & $\textbf{73.7}^{*}$ & 73.3 & 55.8 & 68.6 & $\textbf{70.2}^{*}$ & 70.1 \\ \rowcolor{LightGray}

{$\circ$}  & {$\bullet$} & {$\bullet$} & {$\circ$}  &   15.4 & 63.8 & $\textbf{66.8}^{*}$ & 70.6 & 30.4 & 64.0 & $\textbf{69.7}^{*}$ & 73.9 & 42.6 & 65.3 & $\textbf{67.0}^{*}$ & 71.9 \\ 

{$\bullet$} & {$\bullet$} & {$\circ$}  & {$\circ$}  &71.1 & 83.9 & $\textbf{84.3}$ & 83.3 & 11.9 & $\textbf{56.7}^{*}$ & 55.3 & 54.3 & 1.2 & $\textbf{29.0}^{*}$& 24.2 & 30.7 \\ \rowcolor{LightGray}

{$\circ$}  & {$\bullet$} & {$\circ$}  & {$\bullet$} & 47.3 & 80.8 & $\textbf{82.2}^{*}$ & 83.1 & 17.2 & 53.4 & $\textbf{57.2}^{*}$ & 59.7 & 0.6 & 28.3 & $\textbf{30.7}^{*}$  & 33.4 \\ 

{$\bullet$} & {$\circ$}  & {$\circ$}  & {$\bullet$} & 74.8 & 86.0 & $\textbf{87.5}^{*}$ & 86.3 & 17.7 & 58.7 & $\textbf{59.7}$ & 57.7 & 0.8 & 28.0 & $\textbf{34.6}^{*}$ & 31.0\\ \rowcolor{LightGray}

{$\bullet$} & {$\circ$}  & {$\bullet$} & {$\circ$}  & 68.4 & 83.3 & $\textbf{85.5}^{*}$ & 85.3 & 41.4 & 67.6 & $\textbf{72.9}^{*}$ & 72.0 & 53.8 & 68.0 & $\textbf{70.3}^{*}$ & 69.9 \\ 

{$\bullet$} & {$\bullet$} & {$\bullet$} & {$\circ$}  &  70.2 & 85.1 & $\textbf{86.2}^{*}$ & 85.1 & 48.8 & 70.7 & $\textbf{74.2}^{*}$ & 74.9 & 60.9 & 69.9 & $\textbf{71.1}$ & 70.1 \\ \rowcolor{LightGray}

{$\bullet$} & {$\bullet$} & {$\circ$}  & {$\bullet$} & 75.2 & 87.0 & $\textbf{88.0}^{*}$ & 85.7 & 18.7 & 61.0 & $\textbf{61.5}$ & 57.9 & 1.0 & 33.4 & $\textbf{34.1}$ & 34.1 \\ 

{$\bullet$} & {$\circ$}  & {$\bullet$} & {$\bullet$} & 75.6 & 87.0 & $\textbf{88.6}^{*}$ & 85.8 & 54.9 & 72.2 & $\textbf{75.6}^{*}$ & 75.2 & 60.5 & 69.7 & $\textbf{71.2}^{*}$ & 72.2 \\ \rowcolor{LightGray}

{$\circ$}  & {$\bullet$} & {$\bullet$} & {$\bullet$} & 44.2 & 82.1 & $\textbf{83.3}^{*}$ & 81.5 & 46.6 & 70.7 & $\textbf{75.3}^{*}$ & 74.7 & 55.1 & 69.7 & $\textbf{71.1}^{*}$ & 71.1 \\

{$\bullet$} & {$\bullet$} & {$\bullet$} & {$\bullet$} & 73.8 & 87.6 & $\textbf{88.8}^{*}$ & 87.5 & 55.3 & 73.4 & $\textbf{76.4}^{*}$ & 78.4 & 61.1 & 70.8 & $\textbf{71.7}^{*}$ & 72.7 \\ \hline
\multicolumn{4}{c|}{Means} & 50.7 & 78.6 & $\textbf{80.1}^{*}$ & 81.6 & 28.7 & 59.7 & $\textbf{64.0}^{*}$ & 66.2 & 27.4 & 48.1 & $\textbf{50.0}^{*}$ & 52.7\\\hline
\end{tabular}}
\end{table}

\section{Discussion and conclusion}
In this work, we demonstrate the efficacy of a multi-modal variational approach for segmentation with missing modalities. Our model outperforms the state-of-the-art approach HeMIS \cite{hemis}. In fact, HeMIS could be seen as the non-variational version of our method where: 1/one does not sample but uses the mean of the latent variable instead; 2/the modality-specific covariances are set up to the identity, $\Sigma_i=I$; 3/only the segmentation is reconstructed from the hidden variable. In this case, each modality are independently encoded and averaged such as HeMIS. Finally, our method (\textit{U-HVED}) offers promising insight for leveraging large but incomplete data sets. For future work, we want to provide an analysis of the the learned embedding. This task is particularly challenging due to the multi-scale representation of the hidden variable.

\subsubsection{Acknowledgement}
We thank C. Sudre, W. Li, B. Murray, Z. Eaton-Rosen, F. Bragman, L. Fidon and T. Varsavsky for their useful comments. This work was supported by the Wellcome Trust [203148/Z/16/Z] and EPSRC [NS/A000049/1]. TV is supported by a Medtronic/RAEng Research Chair [RCSRF1819/7/34].

\bibliographystyle{splncs04}
\bibliography{paper1382_camera_ready}

\end{document}